\documentclass[twoside, epsfig]{article}

\input oejv.sty

\usepackage{tablefootnote}
\usepackage{float}

\usepackage{graphicx} % omit 'demo' option in real document
\usepackage[graphicx]{realboxes}

\usepackage{subfig}

\begin{document}

% =========== BEGINNING OF THE PAPER ========================
\OEJVhead{May 2022}
\OEJVtitle{NEW SEMIREGULAR VARIABLE STAR NEAR THE WIZARD NEBULA}
\OEJVtitle{Evolution of the red giant ``MaCoMP\_V1''}
%\OEJVtitle{The red giant MaCoMP\_V1 evolution}
\OEJVauth{G. Conzo$^1$; M. Moriconi$^1$; P.G. Marotta$^1$}

\OEJVinst{Gruppo Astrofili Palidoro, Fiumicino, Italy {\tt \href{mailto:giuseppe.conzo@astrofilipalidoro.it}{giuseppe.conzo@astrofilipalidoro.it}}}

\OEJVabstract{The red giant \textit{\normalsize{MaCoMP\_V1}} in Cepheus at coordinates RA (J2000) \normalsize{$22:49:05.49$} DEC (J2000) \normalsize{$+57:52:41.6$} is a semiregular variable star classified as SRS, number 2225960 in the AAVSO VSX database. Using the Fourier transform, the period \normalsize{$ P = (24.751 \pm 0.062)$\,d} was evaluated and, with the support of the ASAS-SN and ZTF surveys, a well-defined light curve was made.
The analysis resulted in the fundamental physical parameters of \textit{\normalsize{MaCoMP\_V1}}, such as the mass $ M = (4.97 \pm 0.38)$\,M$_\odot$ and radius $ R = (40.5 \pm 6.7)$\,R$_\odot$, with consistent values suggesting the characteristics of a semiregular red giant.
In addition, the effective temperature $ T_{eff} = (4500 \pm 135)$\,K from the $Gaia$ catalog and the stellar evolution based on the \textit{\normalsize{$Sch\ddot{o}nberg$-$Chandraskehar$}} limit was estimated.
}

\begintext

\section{Introduction}\label{secintro}

Semiregular variables are the most widespread stars in the galaxy and are very interesting to study because of their evolutionary complexity. After leaving the Main Sequence, they pass into the Cepheid instability region, transforming into pulsating variables of the $\delta$ Cephei type.

These stars can be described as semiregular giants and supergiant variables with spectral classes ranging from F to K, referred to as SRd. If these variables show high luminosity in the evolutionary process, they fall into the red supergiant region, becoming SRc type. In contrast, they turn into SRab semiregular variables with late spectral classes if they show lower luminosity.
Finally, in the classification of semiregular variables, there are the pulsating low-mass supergiants, which may be in a thermal evolutionary phase such that they move from red giant to protoplanetary nebula \citep{2019Ap.....62..556K}.

A month-long photo session was conducted in 2019 on the Wizard Nebula, an object identified in the open cluster classified NGC\,3780 in the New General Catalog  \citep{1888MmRAS..49....1D}. It is an emission nebula and is known for its ionization due to the presence of the DH Cephei binary star system. Therefore, it has proven to be of great interest in both photographic and photometric studies, in fact the star field is known to have a high presence of giant stars \citep{1994A&A...283..963C}.

Many stars were measured using a 60mm Refractor apochromatic telescope and an ASI-385MM camera. Adopting the Aladin \citep {2011ascl.soft12019C} software, only red stars were chosen and studied because these types have a late evolutionary state, and they are interesting to observe. A star at about 20' from the Wizard Nebula was found through photometric measurements, showing possible brightness variations. To confirm these results, it was necessary to continue observations until the end of 2021 with a better setup and to compare them with automatic surveys using ASAS-SN \citep{2017PASP..129j4502K} and ZTF \citep{2019PASP..131a8003M}.

At the end of January 2022, the American Association of Variable Star Observers (AAVSO) approved the discovery of 2MASS J22490550 +5752417 at 
coordinates RA (J2000) $22:49:05.49$ DEC (J2000) $+57:52:41.6$, and named it \textit{MaCoMP\_V1}. This nomenclature derives from the 
authors: \textit{Ma} (Mara Moriconi), \textit{Co} (Conzo Giuseppe) and \textit{MP} (Marotta Paolo). Finally, \textit{V1} indicates the first 
variable star discovered by the team.
 
\section{Photometric observations}

\begin{figure}[h]
\centering
\includegraphics[width=9cm]{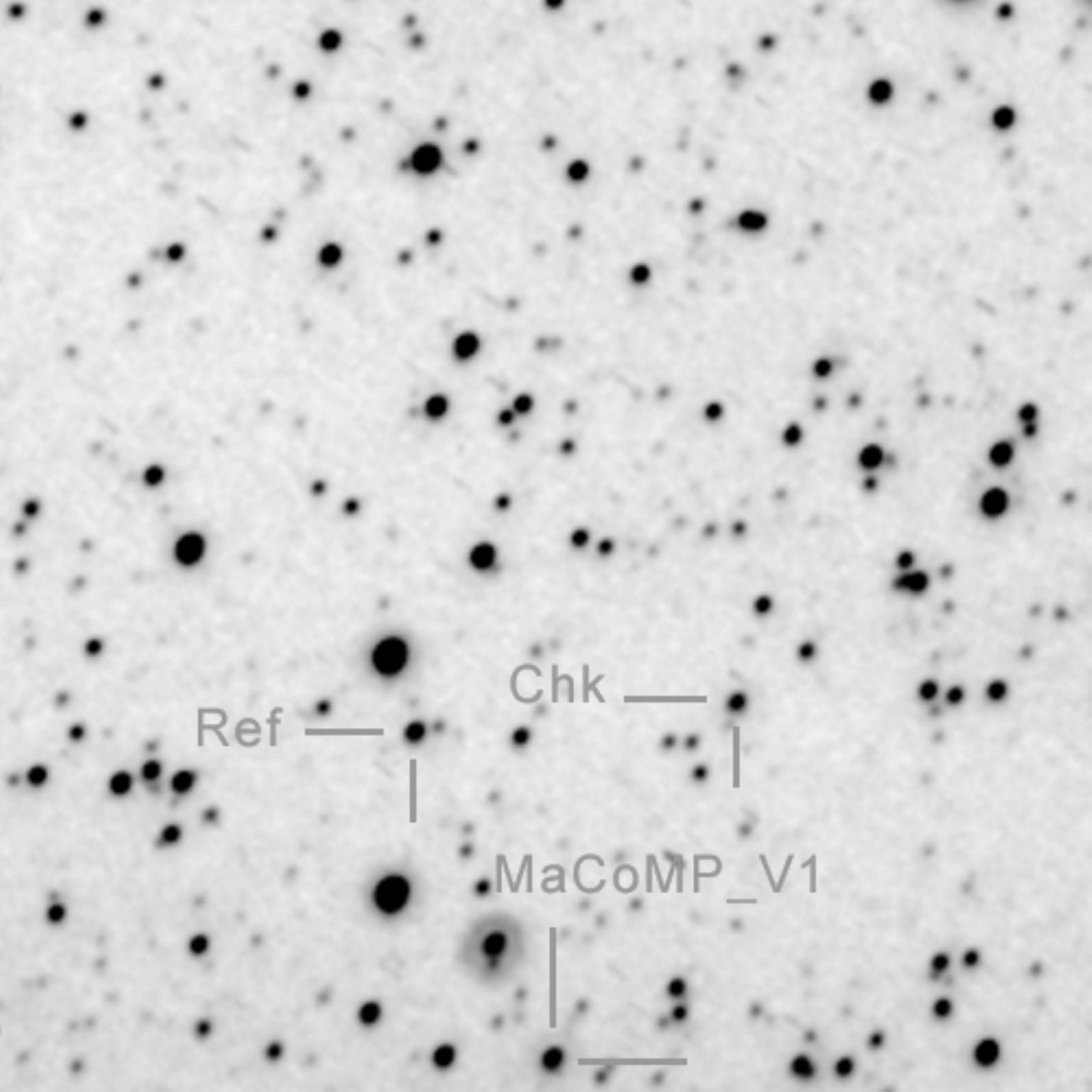}
\caption{Star field with the target, reference, and check stars.}
\label{fig:field}
\end{figure}

The collected images were calibrated appropriately, and photometry was carried out with Maxim DL v6.20 software. The reference and check stars (Check) used for our purposes are listed in Table \ref{tab:stars} and are identified in Figure \ref{fig:field}.

\begin{table}[h]
\centering
\caption{Reference and Check stars used for the photometry.}
\begin{tabular}{llc}
\hline
\hline
Type                 & Name         & Coordinates (J2000)\\
\hline
Reference            & 2MASS J22491561+5754412 & 22 49 15.62 +57 54 41.2\\
Check                & 2MASS J22485978+5755182 & 22 48 59.78 +57 55 18.2\\ 
\hline
\end{tabular}
\label{tab:stars}
\end{table}

The period was obtained with the Fourier Transform, using the Period04 software \citep{2014ascl.soft07009L}, and then the final period was calculated.

\begin{equation}
P = 24.751 \mathrm{d}
\end{equation}

With this value, it was possible to merge the observed data with those from ASAS-SN \citep{2017PASP..129j4502K} and ZTF \citep{2019PASP..131a8003M} surveys. A luminosity variation in cycles of \textit{24 days 18 hours and 14 minutes} identified in the light curve was obtained, the profile of which shows a typical semiregular red giant variation of SRS type \citep{2006SASS...25...47W}.

\begin{figure}[h]
\includegraphics[width=16cm]{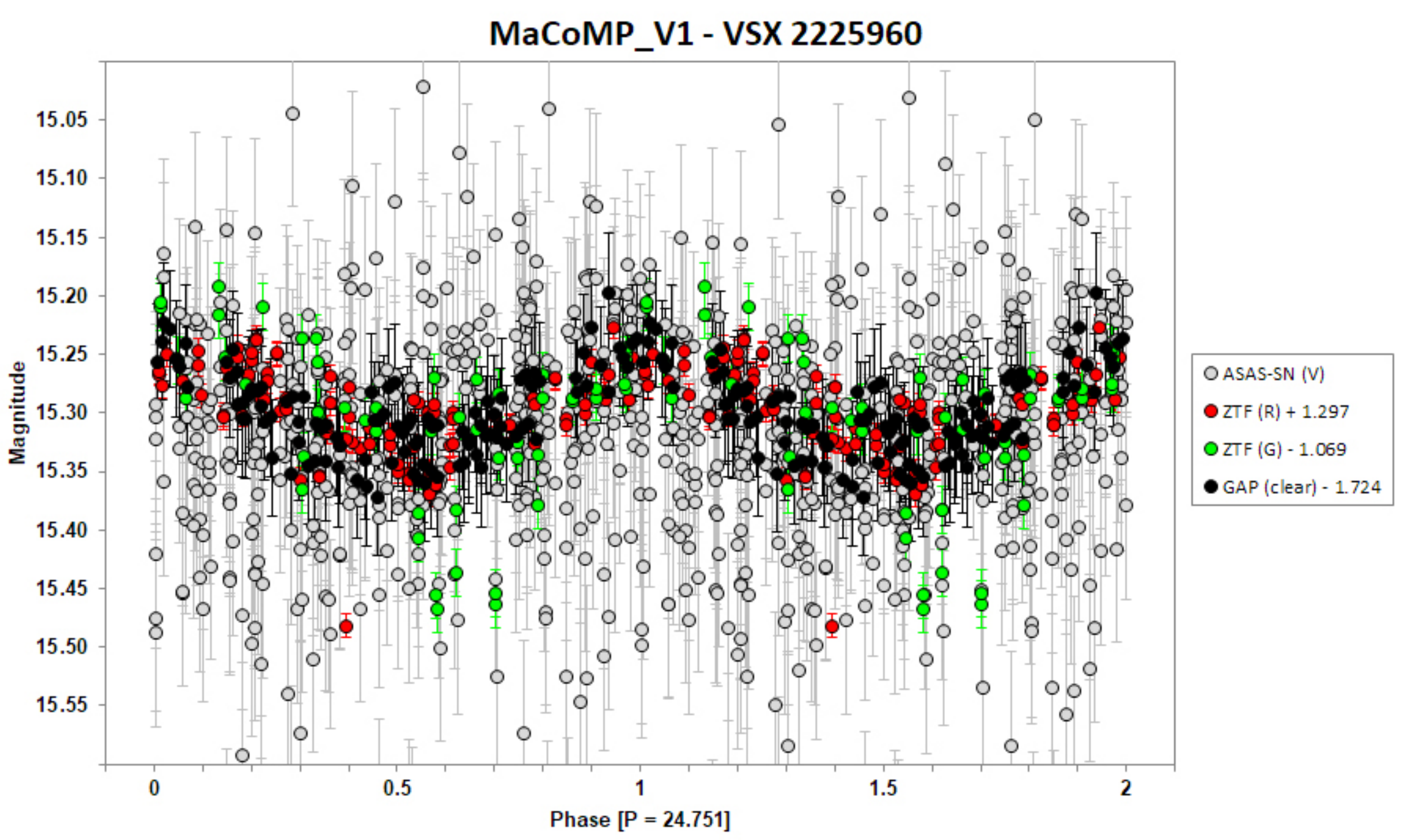}
\caption{Light curve of MaCoMP\_V1, ZTF Survey (Sloan $R$ and $G$ filters) and ASAS-SN Survey (Johnson $V$ filter) compared with GAP data (clear).
\label{fig:plot}}
\end{figure}

The plot shows four sets of data centred on HJD epoch $2458726,641$ at values 0, 1, and 2 in the phase domain when the star is brightest.
The data observed by GAP (black dots in Fig. \ref{fig:plot}) were made in \textit{clear} using two different telescopes:

\begin{itemize}
 \item Apochromatic Refractor Telescope with 600mm focal length and f/7.5 ratio; 
 \item Schmidt Cassegrain Telescope 2000mm focal length and f/10 ratio.
 \end{itemize}

The ZTF survey data (red and green dots in Fig. \ref{fig:plot}) are splitted into the Red Sloan filter \citep{2018JATIS...4a5002B} and the Green Sloan filter \citep{2018JATIS...4a5002B}. The $R$ filter shows an uncertainty $\sigma = 0.01$\,mag, while the $G$ filter, with a larger scatter, shows an uncertainty $\sigma = 0.02$\,mag. The data from the ASAS-SN Survey (gray dots in Fig. \ref{fig:plot}) were made in the Johnson $V$ filter \citep{1990PASP..102.1181B} and have an uncertainty $sigma = 0.08$\,mag.

Since no filter was used during the photometric sessions, it was necessary to adjust the magnitudes to the standard Johnson $V$ magnitude, using the average magnitude of the $Gaia$ catalogue \citep{2021A&A...649A...3R}. The photometric filters have different passbands than the Johnson-Cousins ones, so switching to a different photometric system was necessary.
This was done by adopting the Equ. \ref{eq:conversions} and using the coefficients from the corresponding conversions webpage \citep{2010A&A...523A..48J}.

\begin{equation}
C_{1} = a + b \cdot C_{2} + c \cdot {C_{2}}^2 + d \cdot {C_{1}}^3
\label{eq:conversions}
\end{equation}

With this method, the reference value of stellar magnitude was obtained, defining it as ``zero point'' and expressing all magnitudes in $V$, in fact, the GAP data differ by $1.724$\,mag, the red ZTF data differ by $1.297$\,mag, and the green ZTF data differ by $1.069$\,mag.

\section{Data analysis} \label{sec:stima}

\subsection{Fundamental parameters studying}

From photometric observations, fundamental physical parameters were estimated, starting with the relative magnitude and stellar distance. The absolute magnitude was obtained, using the parallax parameter $ RUWE = 0.971 $ \citep{2020yCat.1350....0G} to adopt a reliable distance from $Gaia$. This resulted in a parallax $ plx = (0.1423 \pm 0.0129)$\,mas and a photometric magnitude $ V = 15.29 $\,mag, which makes it possible to use the magnitude-distance relation \citep{2006JAHH....9..173H}. We considered the extinction $ A_v = (3.29 \ pm 0.16) $\,mag \citep{ext} relative to the direction of the star and the distance $d=(7027 \pm 637) $\,pc from the $Gaia$ parallax, to calculate the absolute magnitude

\begin{equation}\label{abs}
M_{v}=m-5 log(d)+5 -A_{v} = (-2.24 \pm 0.48) \mathrm{mag}
\end{equation}

The brightness of \textit{MaCoMP\_V1} can be estimated, from absolute magnitude and $Gaia$ distance, using the \textit{brightness-magnitude} relationship, like a standard candle at $ 10 $\,pc \citep {un2013}.

\begin{equation}
	L = 10^{\frac{M_{v}-M_{\odot}}{-2.5}} = (609 \pm 185) \mathrm{L_\odot}
\end{equation}

where $M_{v}$ is the absolute magnitude obtained in relation (\ref{abs}) and $M_{\odot}$ is the absolute magnitude of Sun.
Therefore, using $T_{eff}=(4500 \pm 135) $\,K from $Gaia$ EDR3 \citep{2020yCat.1350....0G} and with Stefan-Boltzmann law \citep{sb}, the star radius

\begin{equation}\label{erre}
	R=\sqrt{\frac{L}{4\pi\sigma T_{eff}^4}} = (40.57 \pm 6.67) \mathrm{R_\odot}
\end{equation}

is obtained.
Finally, using the luminosity and the temperature, the mass was calculated \citep{2015AJ....149..131E} as

\begin{equation}
	M=L^{1/4}=T_{eff}\sqrt{R}= (4.96 \pm 0.38) \mathrm{M_\odot}.
\end{equation}

This confirms the correctness of the photometry in relation to the temperature suggested by $Gaia$, but it should be noted that the estimation of the parameters is done analytically. To achieve higher accuracy and veracity, spectrometric observations leading to an estimate of actual temperature and metallicity would be required.

\subsection{\textit{MaCoMP\_V1} evolution based on Sch\"{o}nberg-Chandraskehar limit}

In the case of stars with masses $M> 2$\,M$_{\odot}$
there is a limiting mass value of the core at which accretion must stop before hydrogen is converted into helium. Consequently, the main sequence corresponds to the melting of the initial mass of hydrogen. This can be proved by the \textit{virial theorem} \citep{1978vtsa.book.....C} which defines the surface pressure term which must be less than the maximum value for equilibrium conditions to exist

\begin{equation}
    P_{sup,max} \propto \frac{T_c^4}{\mu_c^4 M_c^2}
\end{equation}

where \textit{c} refers to the stellar core. Stability condition can be written as average weight per particle between the surrounding envelope and the \textit{core}.

\begin{equation}
    \frac{M_c}{M} \le 0.37 \left(\frac{\mu_e}{\mu_c}\right)^2
    \label{collasso}
\end{equation}

where $M$ is the total mass, $\mu_c$ and $\mu_e$ are the average weights for particles in nucleus and envelope which are $\mu_c=1.3$ and $\mu_e=0.6$, respectively. This results in $\simeq 0.10M $ critical mass in solar units.

This procedure is known as the $Sch\ddot{o}nberg$-$Chandraskehar$ limit (SC) \citep{2020MNRAS.499.4832Z} and it studies the collapse of the nucleus on a Kelvin-Helmholtz ($\simeq$ 30 Myr) time scale \citep{2003ApJ...598..560M}. The layers outside the nucleus increase the star size by $\simeq $ 10, transforming it into a red giant.

In the case of \textit{MaCoMP\_V1}, to interpret the relation (\ref{collasso}) it is necessary to find the value of the \textit{core} mass, starting from the star brightness

\begin{equation}
    L \simeq 200L_\odot \left(\frac{M_{core}}{0.3 M_{\odot}}\right) ^{7.6}
    \label{masse}
\end{equation}

Reversing the relationship (\ref{masse}), we get $M_{c} = (0.35\pm0.03) M_\odot$, so by evaluating the mass ratio, it is possible to define whether the stars \textit{core} is close to instability.

\begin{equation}
     \frac{M_c}{M} = 0.070 \pm 0.008 
     \label{sx}
\end{equation}   
\begin{equation}
     0.37 \left(\frac{\mu_e}{\mu_c}\right)^2 = 0.079 \pm 0.007
     \label{dx}
\end{equation}

The result of the relation (\ref{sx}) is smaller than the result of the relation (\ref{dx}), then \textit{MaCoMP\_V1} meets the mass limit value defined in the law (\ref{collasso}), ensuring the stability of the stellar \textit{core}.
\textit{MaCoMP\_V1}, as a red giant star, is very bright and, according to stellar evolution, has completed hydrogen-burning but has not yet collapsed. It has not finished nuclear reactions within the core, and helium burning has begun
\citep{2020MNRAS.499.4832Z}.

\section{Results}

Observations conducted near the Wizard Nebula between 2019 and 2021 revealed a suspected change in the star's brightness. Photometry sessions led to define that the star 2MASS J22490550 +5752417 at coordinates RA (J2000) $22:49:05.49$ DEC (J2000) $+57:52:41.6$ is an SR-type variable \citep{2019Ap.....62..556K}. After data collection, it was approved and registered under VSX code 2225960 in the Variable Star Database \citep{2015yCat....102027W}.

\textit{MaCoMP\_V1} was shown to be a \textit{semiregular red giant} with brightness variation over the period $ P = 24.751 $\,d, describing typical pulsating stars trend shown in Figure \ref{fig:plot}.
The variability, the star stability and, according to the SC limit, \textit{core} stability was analysed.
Data analysis led to the physical characteristics defined in Table \ref{tab:colori}.

\begin{table}[h]
\centering
\caption{Fundamental parameters of \textit{MaCoMP\_V1}}
\begin{tabular}{ll}
\hline
\hline
Name                 & MaCoMP\_V1\\
\hline
Variability type     & SRS \\
Period               & 24.751 $\pm$ 0.062\,d\\ 
Distance             &  7027 $\pm$ 637\,pc\\
Absolute Magnitude   & -2.24 $\pm$ 0.48\,mag\\
\hline
Luminosity           &  609 $\pm$ 185\,L$_{\odot}$\\
Temperature          &  4500 $\pm$ 1385\,K\\
Radius               &   40.57 $\pm$ 6.67\,R$_{\odot}$\\
Mass                 &   4.69 $\pm$ 0.38\,M$_{\odot}$\\
\hline
\end{tabular}
\label{tab:colori}
\end{table}

\begin{figure}[h]
\centering
\includegraphics[width=12cm]{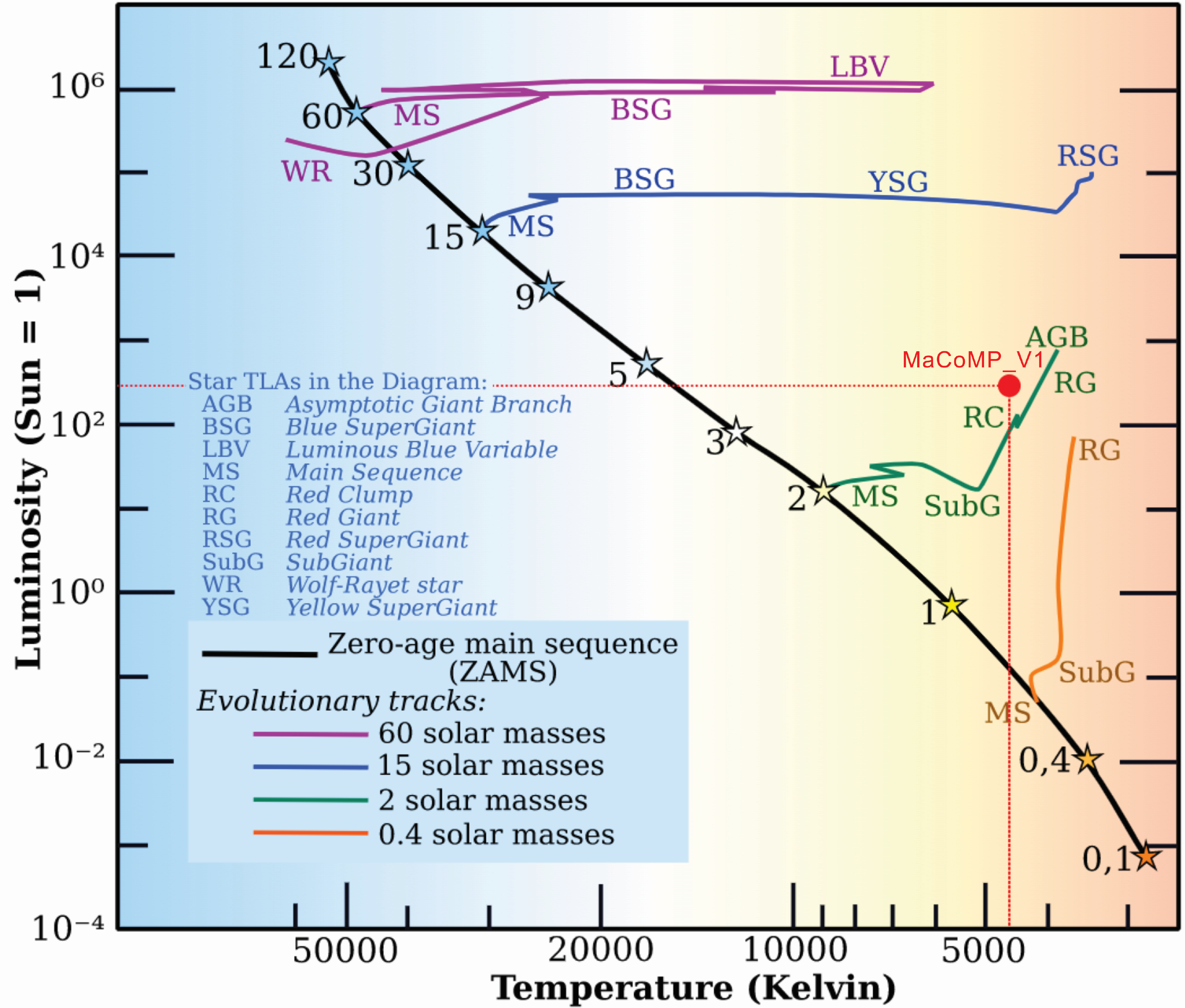}
\caption{\textit{MaCoMP\_V1} (red circle) in stellar evolutionary tracks diagram \citep{ev}.}
\label{fig:hr}
\end{figure}

Evolutionary tracks are defined by the physical processes that take place inside stars.
Objects with different masses will have different evolutionary paths \citep{sev}.
\textit{MaCoMP\_V1} has more than two solar masses, so it should have left the Main Sequence and should arrive in the Red Giant branch. The found parameters allow to place the star in the diagram shown in Figure \ref{fig:hr}.

%\vskip 0.5cm

\end{document}